\documentclass{IEEEtran}
\usepackage{cite}
\usepackage{amsmath,amssymb,amsfonts}
\usepackage{graphicx}
\usepackage{textcomp,nicefrac}
\usepackage{url}

\def\BibTeX{{\rm B\kern-.05em{\sc i\kern-.025em b}\kern-.08em
T\kern-.1667em\lower.7ex\hbox{E}\kern-.125emX}}
\begin{document}
\title{Data taking network for COMET Phase-I}

\author{Youichi Igarashi and Hiroshi Sendai


\thanks{
Manuscript received Oct 30, 2020.
}
\thanks{
This work was supported by JSPS KAKENHI Grant Numbers 17H06135 and 20K04005.
}
\thanks{
Y.~Igarashi is with Institute of Particle and Nuclear Studies,
J-PARC Center,
High Energy Accelerator Research Orgnization,
Tokai-mura, Ibaraki, Japan
(e-mail: youichi.igarashi@kek.jp).
}
\thanks{
H.~Sendai is with Institute of Particle and Nuclear Studies,
J-PARC Center,
High Energy Accelerator Research Orgnization,
Tokai-mura, Ibaraki, Japan
(e-mail: hiroshi.sendai@kek.jp).
}

}

\maketitle

\begin{abstract}
An experiment to search for mu-e conversion named COMET is being constructed at J-PARC. 
The experiment will be carried out using a two-stage approach of Phase-I and Phase-II.
The data taking system of Phase-I is developed based on common network technology.
The data taking system consists of two kinds of networks.
One is a front-end network.
Its network bundles around twenty front-end devices that have a 1~Gb optical network port.
And a front-end computer accepts data from the devices via its network.
The other is a back-end network that collects all event fragments from the front-end computers using a 10~Gb network.
We used a low price 1~Gb~/~10~Gb optical network switch for the front-end network.
And direct connection between the front-end PC and an event building PC using 10~Gb optical network devices was used for the back-end network.
The event building PC has ten 10~Gb network ports.
And each network port of the event building PC is connected to the front-end PC's port without using a network switch.
We evaluated data taking performance with an event building on these two kinds of networks.
The event building throughput of the front-end network achieved 337~MiB/s.
And the event building throughput of the back-end networks achieved 1.2~GiB/s.
It means that we could reduce the construction cost of the data taking network using this structure without deteriorating performance.
Moreover, we evaluated the writing speed of the local storage RAID disk system connected to a back-end PC by a SAS interface, and a long-distance network copy from the experiment location to the lasting storage.
\end{abstract}


\section{Introduction}
\label{sec:introduction}
\IEEEPARstart{A}{n} experiment to search for coherent neutrino-less conversion of a muon to an electron in a muonic atom ($\mu-e$ conversion) named COMET is now constructed in the Hadron hall.\cite{comet}
The experiment tries to find a process $\mu^{-}N \rightarrow e^{-}N$ to search for charged lepton flavor violation (LFV). 
This LFV process is a clear signal of new physics beyond the Standard Model.
The experiment will run in two stages, Phase-I and Phase-II.
Phase-I aims for a signal sensitivity of O($10^{-15}$). 
And we will try to achieve a signal sensitivity of O($10^{-17}$) in Phase-II.
We will discuss the data-taking network in the Phase-I. 
The two kinds of central systems are used in the detector system of the Phase-I measurement.
The one is a straw tracker and an electromagnetic calorimeter by LYZO crystals.
The first system measures the muon beam.
The other is a Cylindrical Drift Chamber (CDC) and Cherenkov trigger counters.
The second system detects an electron from a muonic atom and measures the physics observables.

We adopted the "On-detector readout" concept to improve the signal to noise ratio and reduce the number of cables from the detectors.
Hence, the readouts of the detectors are mounted on the detectors in the magnetic field of 1~Tesra.
Due to this requirement, the detector's readouts are equipped with  1~Gbps optical Ethernet ports to send their data.
A data acquisition (DAQ) system collects data from these detector systems via networks.
The final destination of acquired data is a  mass storage system in KEK Computer Research Center, located 60~km, far from the J-PARC Hadron hall.
Therefore, the acquired data are stored once in local storage then sent to the computer research center as a bucket-brigade.

Although the estimation of the data rate is not exact in the current stage, the value of 300~MiB/s from many assumptions is used as a guiding value of the performance.

To achieve this guiding value, we had planned to use a high-performance network switch, e.g., products of Cisco Systems, Inc.
However, it would raise detector construction costs.
On the other hand, several new companies started to deliver low intelligence but low prices network switches.
These switches have attractive specifications because the data taking network does not need intelligent functions on Layer~3.
However, we are worried about performance and specifications that weren't written in their datasheet.
Network-based event building requires a data buffer with sufficient size, internal process speeds, and internal bandwidth.
In order to investigate whether the low-price network switch can be acceptable or not, we evaluated an actual data taking network configured by the low-price network switch.
Likewise, a more high-performance network switch would have been required for the back-end side on which the entire data focus.
In order to suppress the assembly cost, we configured the back-end, instead, as a point to point connection by 10GBASE-SR Ethernet.
And we have evaluated whether the direct connection network works.
Moreover, we evaluated the performance of local storage and the speed of data transfer to permanent storage.
Based on these, we estimated the entire performance of the whole data taking.

In this article, the structure of the COMET Phase-I data-taking network, the several evaluations of these networks, and the expectation of the entire performance of the data-taking system are reported.
\section{COMET Phase-I data taking network}

\subsection{Design of the data taking network of the COMET experiment}

\begin{figure*}[bth]
\centerline{\includegraphics[width=4.5in]{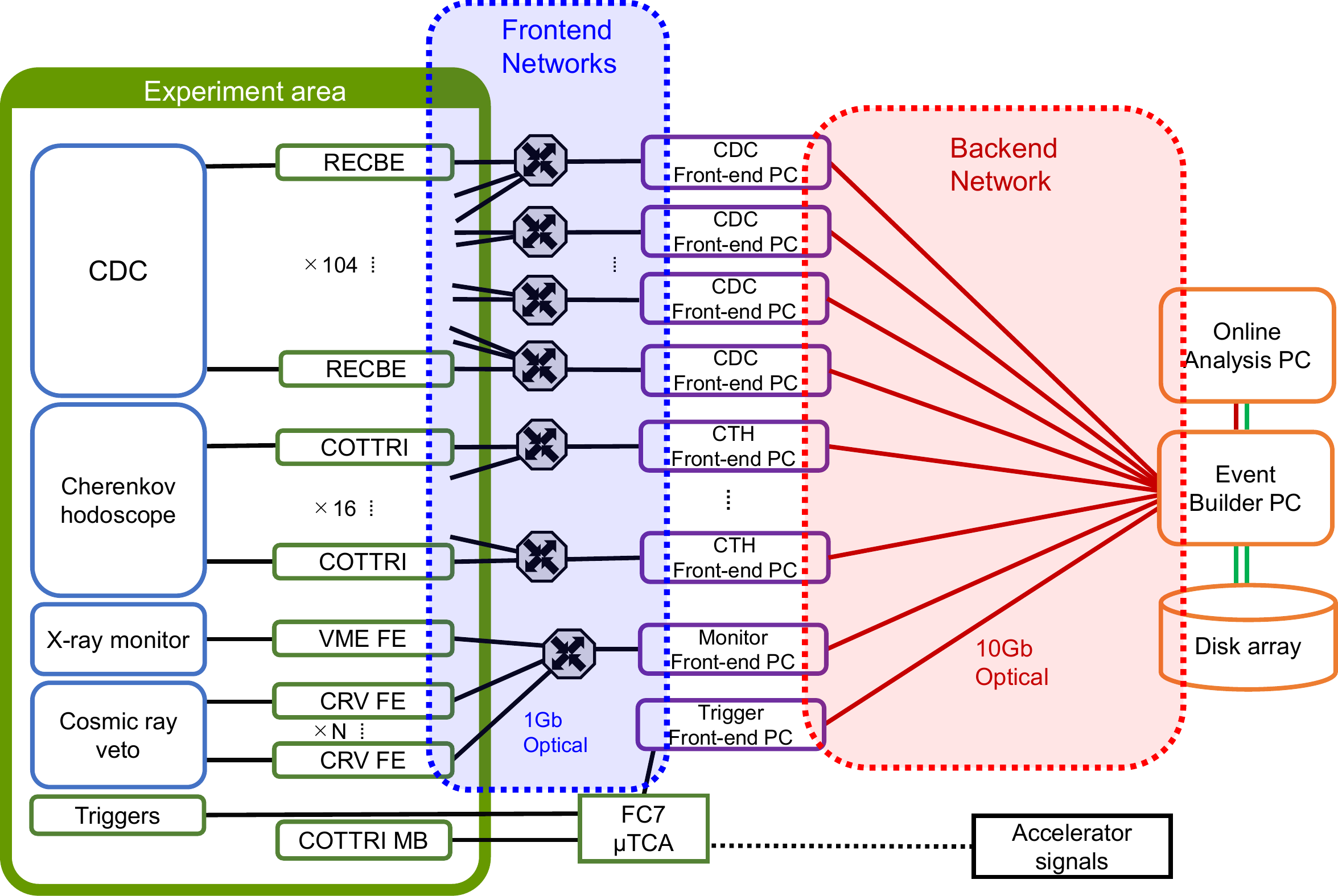}}
\caption{The conceptual design of the COMET Phase-I DAQ network}
\label{fig:daqnet}
\end{figure*}

The DAQ system of the COMET detector system is based on the generic TCP/IP network technology. 
The number of front-end readout devices are more than 150.
The readout network consists of a two-layer structure to handle front-end devices of such large numbers.
One of the layers is a front-end network, and the other is a back-end network.
The configuration of the DAQ network is shown in Fig.~\ref{fig:daqnet}.

The front-end network is located in front of each front-end PC. And the PC collects data from around twenty readout devices.
The front-end PC sends the collected data to a back-end PC via the back-end network.
We have found a cost-effective solution for each two sub-network while keeping data transfer performance.

\subsection{Front-end network}

\begin{figure}[htb]
\begin{center}
\includegraphics[width=3.0in]{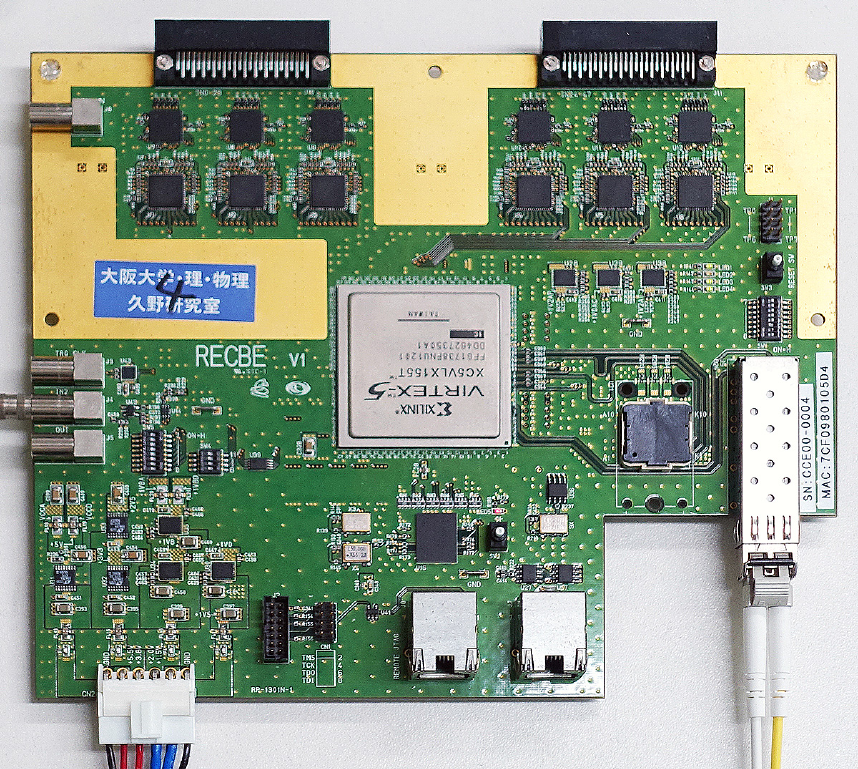}
\caption{
RECBE; A readout card of the CDC which has ASD, FADC and TDC.
The readout port uses an optical network link by an SFP connector.
}
\label{fig:recbe}
\end{center}
\end{figure}

The readout of the COMET Phase-I detector adopted the "On detector readout" concept.
Readout devices that have amplifiers, analog to digital converters and data transfer ports are mounted on the detector directly.
They should work well in the 1~Tesla magnet field inside the solenoid. 
Each readout device connects to the front-end PC by the standard TCP/IP network using an FPGA based TCP/IP engine SiTCP\cite{sitcp}.
The metal cable Ethernet did not work under our condition because it uses a pulse transformer that does not work in the magnet field.
Therefore, readout devices use the optical network ports to send their data.
Figure.\ref{fig:recbe} shows a readout board of the COMET CDC. 
It has amplifier discriminator ASICs, analog to digital converters, a controller FPGA, and an SFP interface that handles the 1000BASE-SX optical network.

\begin{figure}
\begin{center}
\includegraphics[width=3.2in]{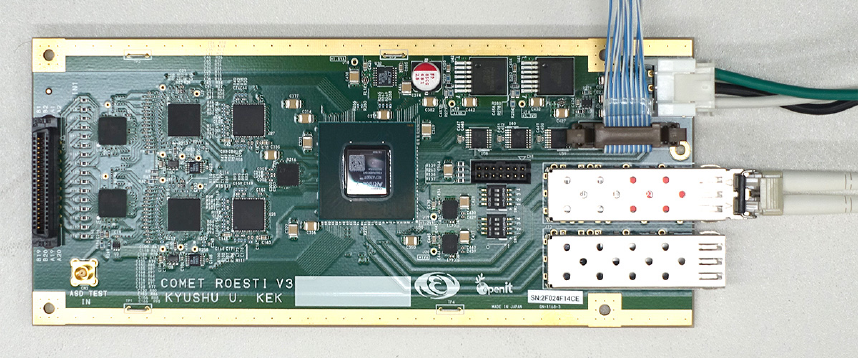}
\caption{ROESTI; A readout card of the straw chamber which has DRS4 based waveform sampler.
The card has two SFP connectors, and transport data by the daisy chain connection using them.}
\label{fig:roesti}
\end{center}
\end{figure}

Fig.\ref{fig:roesti} shows a readout board of the straw tracker with a capacitor-array-based waveform sampler ASIC DRS4\cite{drs4}, a controller  FPGA, and also two SFP interfaces. 
Two readout interfaces mean that the board supports the Daisy-chain function on the TCP/IP network. Multiple boards can be read by the terminated port.

The front-end PC reads over twenty optical network links.
Thus, a 1G/10G high-speed optical network switch is required in front of the front-end PC.
The DAQ system needs to read around 200 optical links.
Therefore such an optical network switch was costly.
However, several companies deliver cost destructive optical network switches nowadays.
FS.COM\cite{fs.com} S3900 24F4S is one of these.
It has twenty 1~Gbps SFP ports, four 1~Gbps Metal ports, and four 10~Gbps SFP+ ports.
And it handles Layer2+ functions.
Its cost is around 1/30 compared with the previous high-performance 1G/10G network switches, and its specification is fit for the requirements of the experiment. 
However, the undocumented specification, such as the buffer size, is not clear.
And we were not sure whether the network switch could be used in the event building. 
We have evaluated it by the actual data taking.

\subsection{Back-end network}

The back-end network combines the front-end PCs and the back-end PC, and it handles whole data of all detectors.
A high-performance 10G network switch is expected to have enough performance to treat the whole data.
On the other hand, a 2U server PC has many high-speed PCI-e slots.
An 8-lane Gen3 PCI-e has 7.9~GB/s transfer speeds and connects directory into the processor inside in the case of the current Intel processor.
This structure means that several multiple-port network cards can transfer data to the processor memory simultaneously.
Therefore the direct connection of the front-end PCs and the back-end PC is expected to work with good performance.
Thus, We adopted the direct connection network without a high-performance network switch to achieve good performance and cost reduction.

\subsection{Local storage}

The DAQ system has local storage to keep the data from detectors for at least several days.
The data is transferred from there to KEK Computer Research Center's data storage simultaneously during the experiment running.

Several high-speed interfaces for data storage are available for general usage; Fiver channel, iSCSI on 10Gbps ethernet, and SAS (Serial Attached SCSI).
We adopted SAS 3.0 for the local data storage.
The transfer speed of SAS 3.0 is 12 Gbps and 1200 MB/s effectively.
It has enough throughput to meet our guiding throughput value of 300~MiB/s, even considering a safety factor of three times.
A RAID disk system with RAID level 5+0 is used for the data storage.
The disk system has forty-two HDD slots, and ten 7200~rpm~/~10~TB HDDs are installed there.
The HDDs can be extended if necessary.
The HDDs are configured as RAID level 5+0 to achieve high throughput and failsafe performance.

\subsection{Long distance data transport}

The final destination of the experiment data is a large scale tape mass storage in the KEK Computer Research Center located in Tsukuba city, which is 60~km away from J-PARC on Tokai village.
This fact means the lasting storage system locates around 60~km away from the experimental location.
The experiment data is stored in the local storage by the DAQ at once. 
The data are then transferred to the mass data storage in the distance, similar to a bucket brigade, via the wide-area network for science and education, named Science Information NETwork (SINET).
J-PARC and KEK keep 10~Gbps bandwidth between them on SINET.
The local storage can keep experiment data for several days to provide for troubles of wide-area communication.
\section{Performance evaluation of each network}

\subsection{Performance evaluation of the front-end network}
\begin{figure}[h]
    \begin{center}
      \includegraphics[width=3.0in]{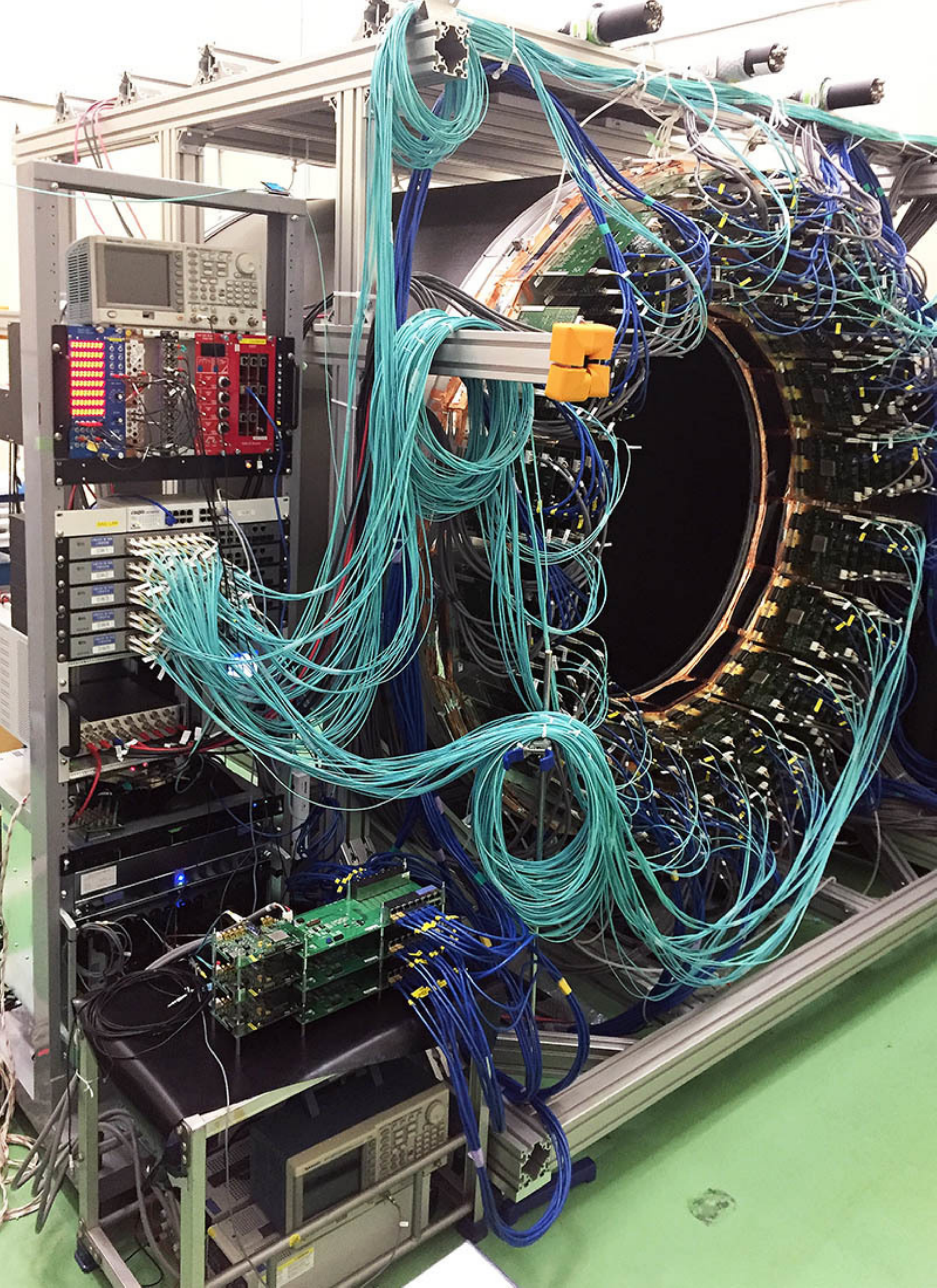}
      \caption{Cosimic ray test setup of the COMET CDC deterctor.}
      \label{fig:cdc}
    \end{center}
\end{figure}

\begin{figure}[h]
\centerline{\includegraphics[width=3.5in]{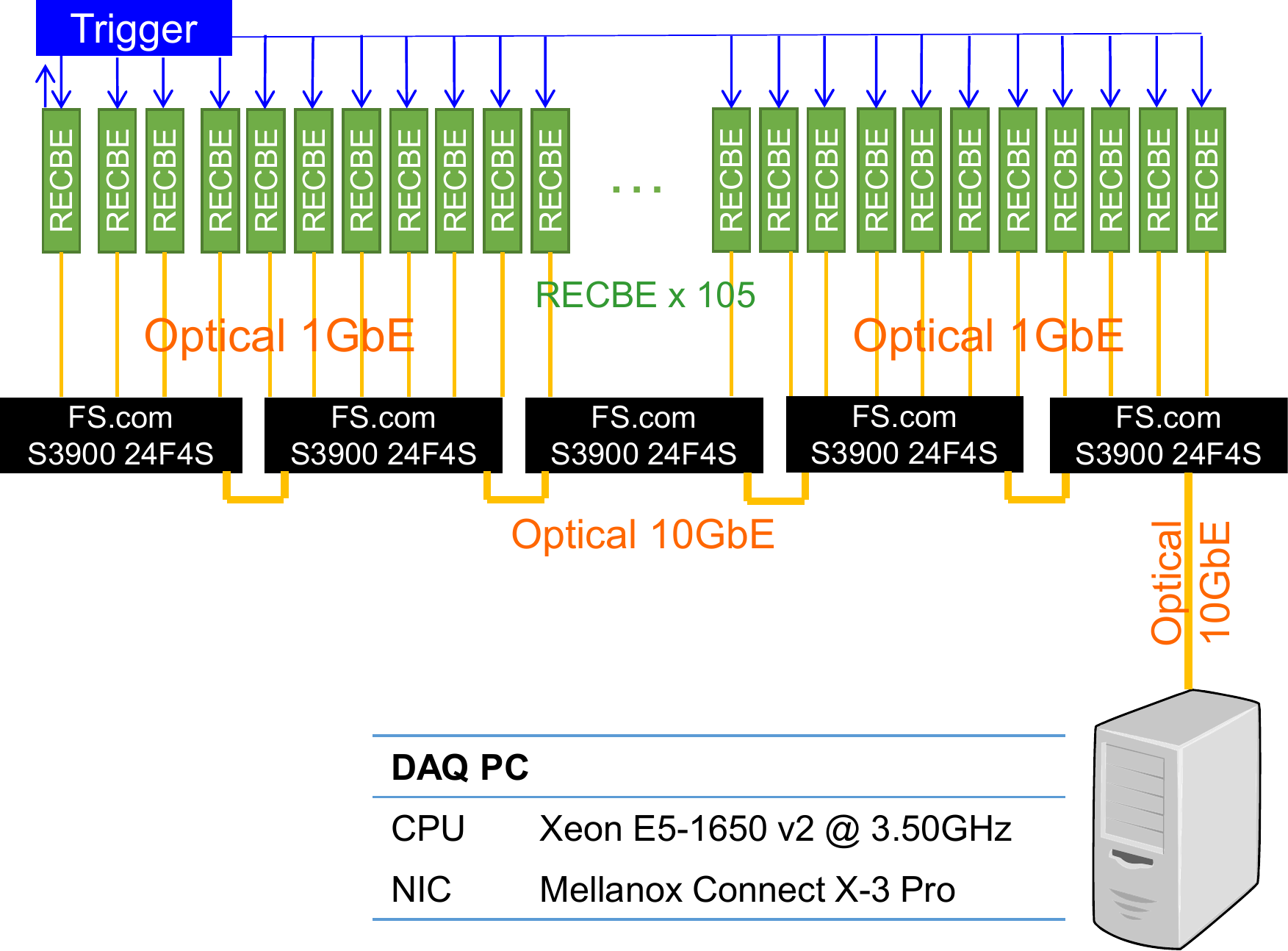}}
\caption{The readout configuration of the COMET CDC test setup using cosmic rays}
\label{fig:cdccrtdaq}
\end{figure}

The evaluation keypoint of the front-end network is whether the low-cost network switch can be usable for the event building or not.
A test of the COMET CDC detector using the cosmic ray is ongoing.
This CDC setup is suitable for evaluating the network switch and then used for the evaluation.
The 104 front-end readout electronics (RECBE) were already installed on it.
One more RECBE that reads the trigger counters is used for trigger generation.
Therefore, a total of 105 RECBEs were working on the setup.
The part of the trigger system was connected to the setup to distribute triggers to RECBEs. 
Figure~\ref{fig:cdc} shows the picture of the CDC cosmic ray test setup.
This CDC test setup is ideal for the evaluation of the network switch.
Therefore we installed the five FS.COM's network switches to it.
The PC with 3.5 GHz Xeon E5-1650v2 and Mellanox Connect X-3 Pro network interface card read all 105 RECBEs data via the network switch.
The operating system CentOS~7.7 worked on the PC for the test.
The configuration of the setup is shown in Fig.~\ref{fig:cdccrtdaq}.

The trigger for the evaluation is generated by a function generator that generates frequent pulses.
The setup did not handle the BUSY signal from each RECBE board.
RECBE has two modes.
One is a RAW mode that outputs 6144~Bytes fixed-length TDC and ADC data.
The other is a SUPPRESS mode that outputs only hit channels TDC and ADC data.
Therefore, the size of the SUPPRESS mode data is variable.
The data come from white noises from the amplifiers in the test.
The size of the data is tuned as around 380~Bytes by the threshold level of the signal height.

\begin{figure}[h]
\centerline{\includegraphics[width=3.5in]{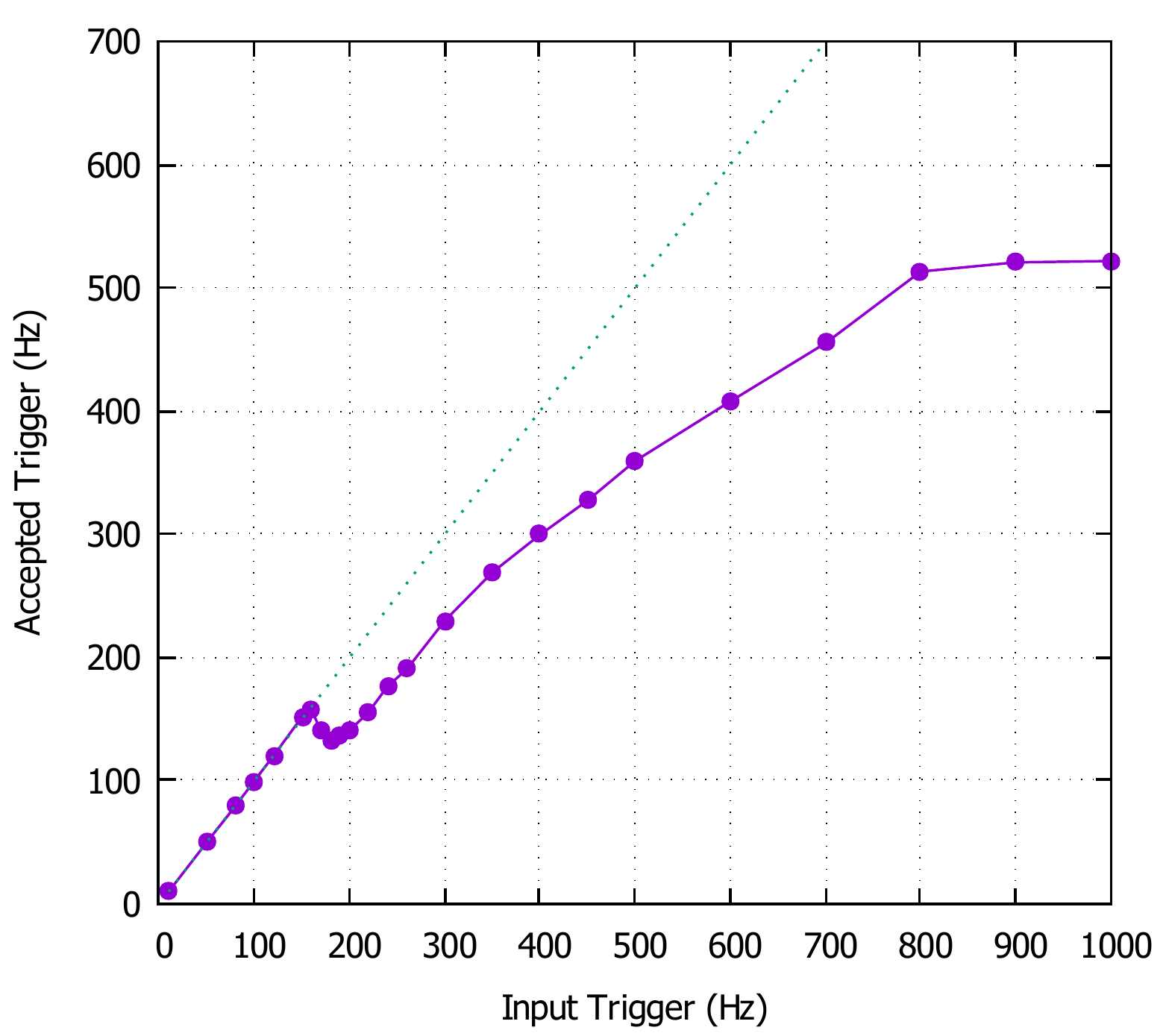}}
\caption{
	Event building performance of the network from 105 RECBE cards
	using 5 FS.COM network switchs and a PC with Xeon E5-1650v2:
	The horizontal axis is the input trigger frequency and the vertical axis is the throughput of the event building.
	}
\label{fig:recbe105_raw}
\end{figure}

Figure.\ref{fig:recbe105_raw} shows the accepted trigger rate using the RAW mode.
The maximum accepted trigger frequency achieved 800~Hz, and the throughput of the data at the maximum trigger rate is 337~MiB/s.
The processor performance of event building limited this accepted trigger rate.
We guess that the reduced performance around 200~Hz is caused by the interference between the buffer size and the frequency.

\begin{figure}[h]
\centerline{\includegraphics[width=3.5in]{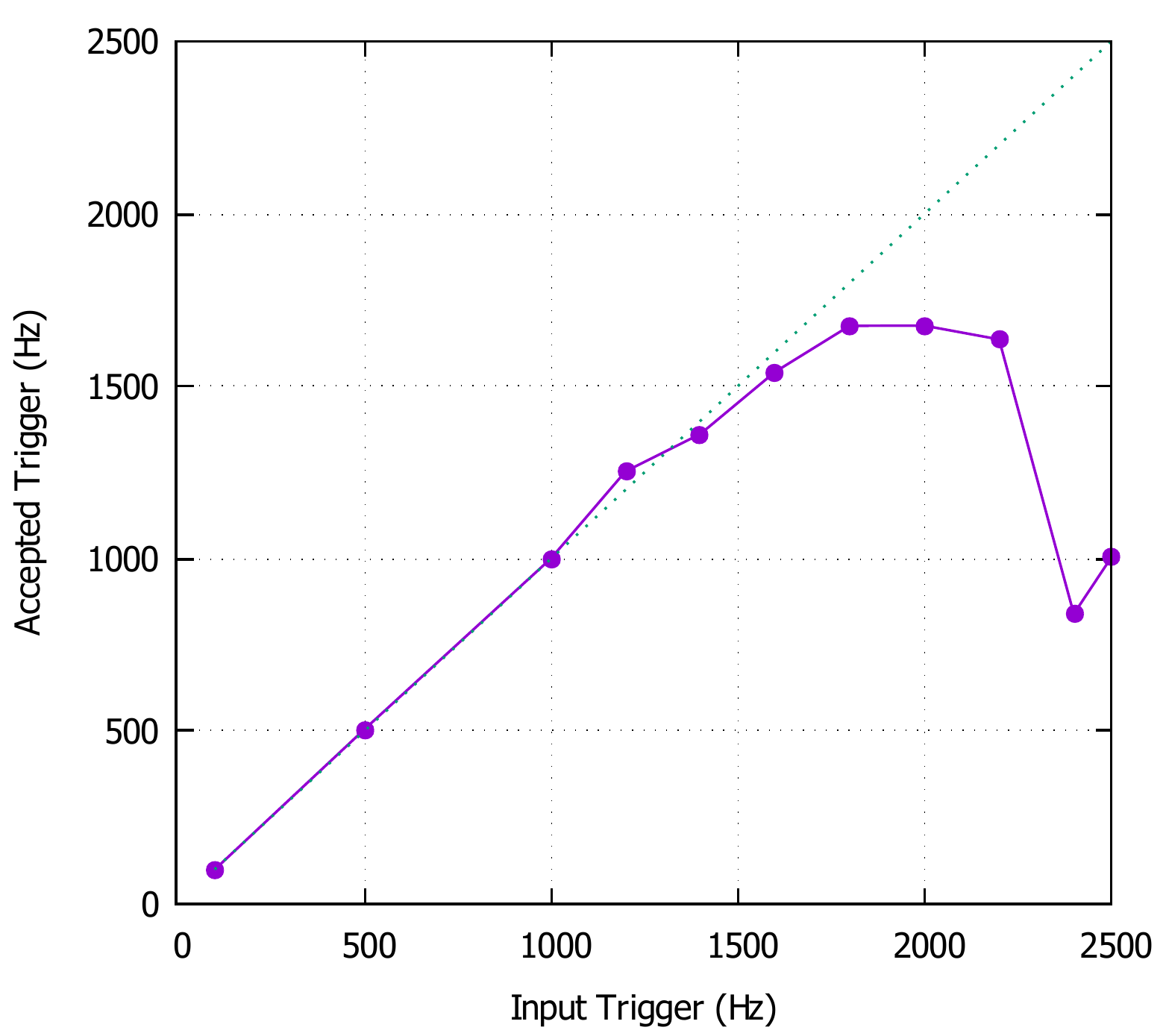}}
\caption{A network event building performance using RECBE suppress mode.}
\label{fig:recbe105_suppress}
\end{figure}

Figure.\ref{fig:recbe105_suppress} shows the accepted trigger rate using the SUPRESS mode.
The accepted trigger rate was saturated at 1800~Hz, and the throughput of the data is 24.4~MiB/s.
When the trigger rate was faster than that, the event building started to fail.
The success rate of the event building is going down to zero after several minutes.
The event building without the BUSY handling caused this slowdown.
However, the low price network switch could handle over 300~MiB/s data from 105 sources.

\subsection{Performance evaluation of the back-end network}

The back-end network was evaluated using the actual data taking PCs.
Two event-building PCs and four front-end PCs are installed in the computer room in the experimental hall.
Those event-building PCs had the same specification, one for the event building and the other as a backup in case of trouble, and it is used for the data copy.
Furthermore, several more PCs are used for the data taking of the experiment.
The specifications of DAQ PCs are summarized in Table.~\ref{tbl:daqpc}.
Scientific Linux 7.9 was used for the operating system of the data taking PCs.

\begin{table*}[htb]
\begin{center}
\caption{Specifications of the DAQ PCs}
\begin{tabular}{llllll}
	PC
	& Processor
	& Memory
	& 10~Gbps NIC
	& SAS/RAID system
	& HDD \\
\hline \hline
	EB     
	& \begin{tabular}{l} Xeon Gold 6126 \\ @ 2.6GHz \end{tabular}
	& 64 GB
	& \begin{tabular}{l} Intel \\ Network Adapter \\ X710-DA4 \end{tabular}
	& \begin{tabular}{l} Broadcom/LSI \\ MegaRAID SAS-3 \\ 3108 \end{tabular}
	& \begin{tabular}{l} SEAGATE 10~TB \\ 7200RPM \end{tabular} \\
\hline
	FE
	& \begin{tabular}{l} Xeon E-2134 \\ @ 3.5GHz \end{tabular}
	& 32 GB
	& \begin{tabular}{l} Mellanox \\ Connect-X 3 Pro \end{tabular}
	&
	&
\end{tabular}
\label{tbl:daqpc}
\end{center}
\end{table*}

\begin{figure}[h]
  \begin{center}
    \includegraphics[width=3.5in]{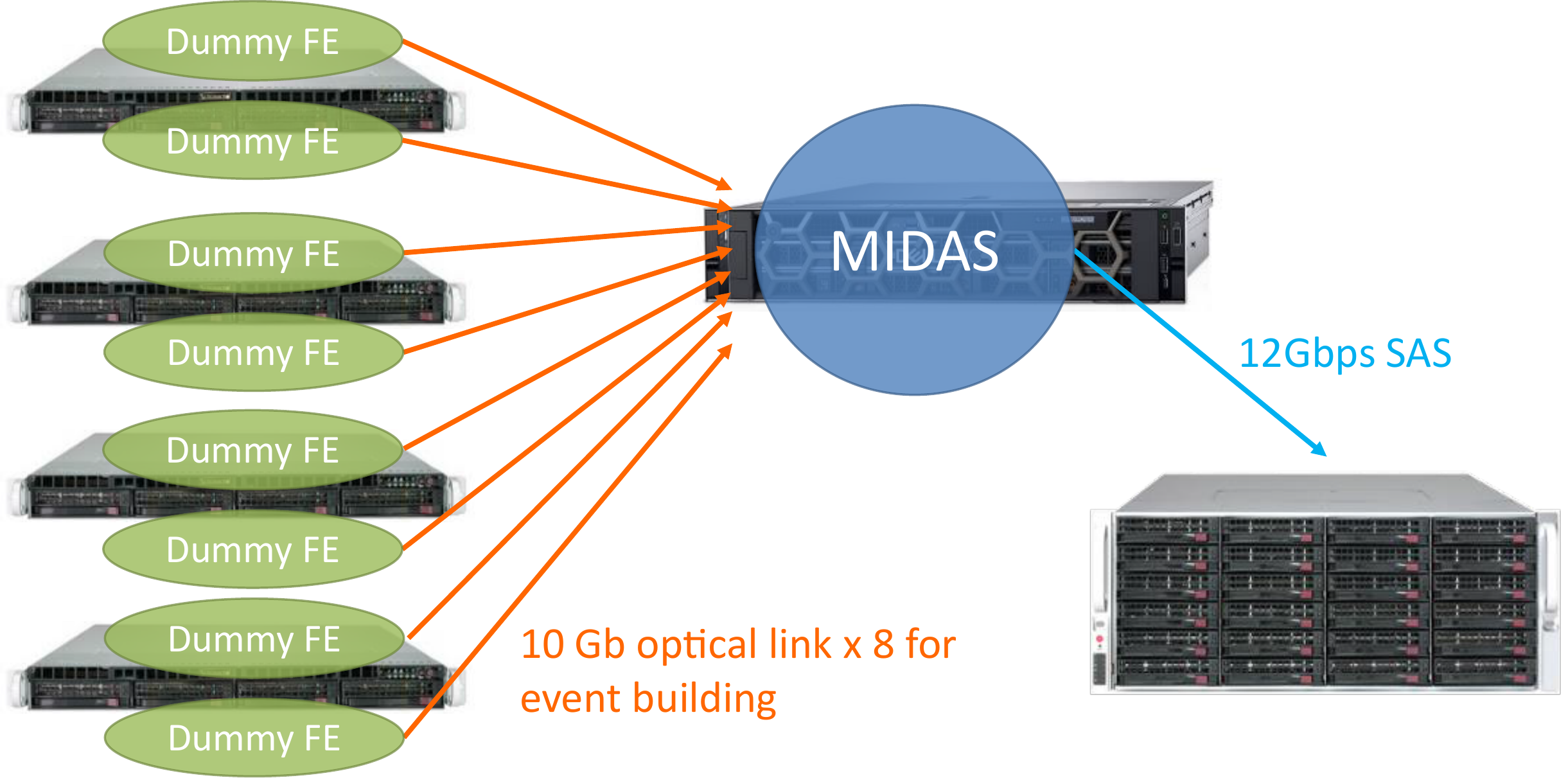}
    \caption{A configuration of a measurement of the event building.
	  A back-end PC has two four-port NICs and two onboard NICs.
	  Four front-end PCs have four-port NICs.}
    \label{fig:testsetup}
    \end{center}
\end{figure}

The front-end PC has two 10~Gbps 10GBASE-SR ports.
One is for the front-end network, and the other is for the back-end network.
Both two network ports are connected to the event building PC to evaluate the back-end network.
The event building PC can receive up to eight connections using four front-end PCs for the event building test.
The event building PC has a RAID disk system connected to the 12~Gbps SAS~(Serial Attached SCSI) interface.
The configuration of the back-end network evaluation is shown in Fig.~\ref{fig:testsetup}.

Two dummy front-end process is running on each front-end PC.
The program sends a 16~kiB dummy event filled with random numbers.
A DAQ software MIDAS\cite{midas} is used for the back-end data taking.
Data-reading processes corresponding to the front-end connections, an event building process, and a data recording process run on the event building PC.
The event building PC has enough processor cores. 
Therefore, each process can use the full performance of a processor core.

\begin{figure}[h]
\centerline{\includegraphics[width=3.5in]{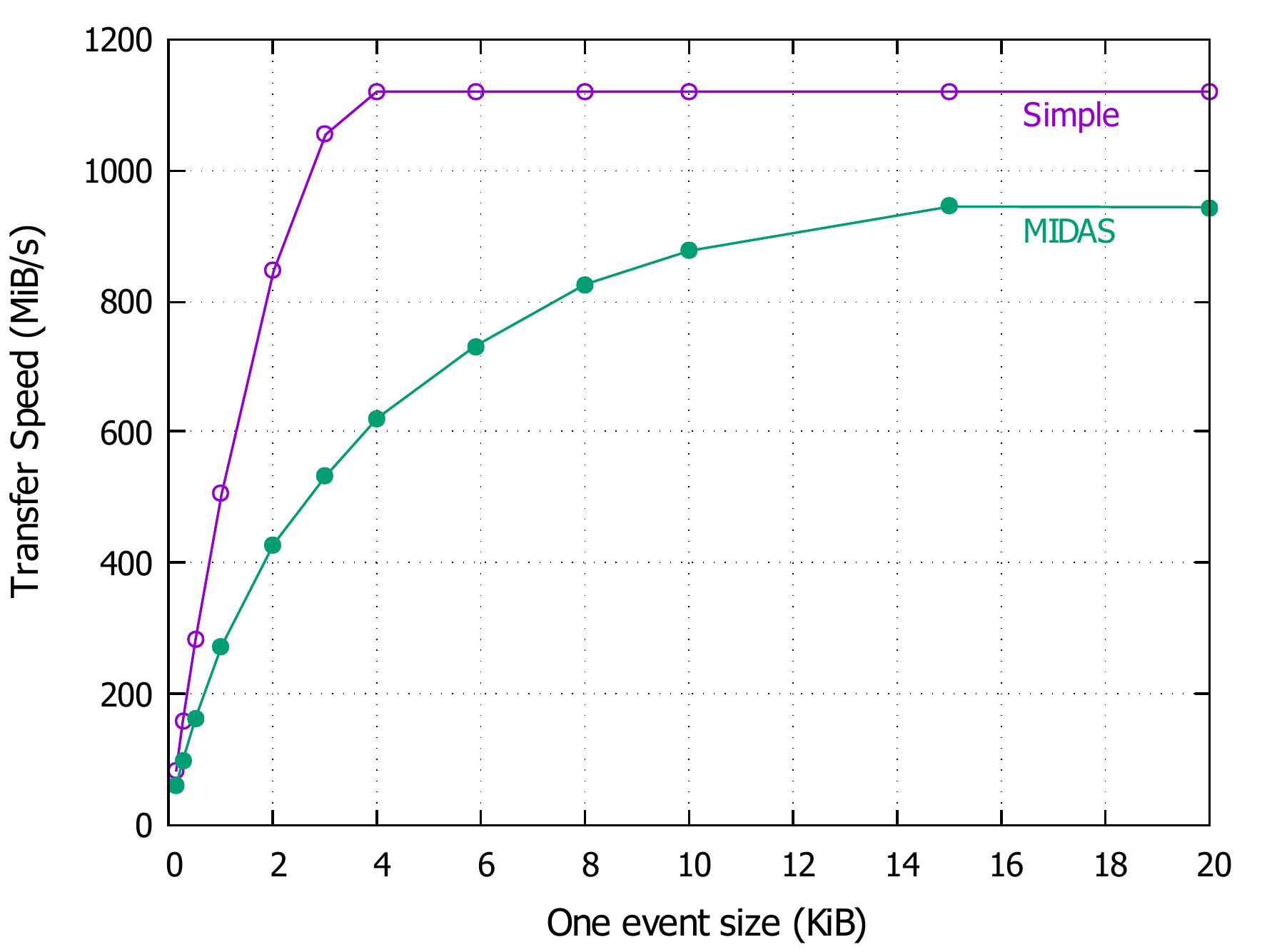}}
\caption{The data size dependency of the data trasfer speed between the front-end PC and the back-end PC.
Open dots show the speed using a simple TCP/IP data transfer program. 
Filled dots show the speed using MIDAS.
}
\label{fig:netspeed}
\end{figure}

Figure~\ref{fig:netspeed} shows the simple data transfer speed of one connection using TCP/IP.
The transfer speed achieved 1.1~GiB/s using an over 4~kiB event size at the simple TCP/IP data transfer.
In the MIDAS case, the data transfer has offset times from the data buffer allocation and arbitration.
Therefore, the event size of over 16~kiB is required to reach the maximum transfer speed.

\begin{figure}[h]
\centerline{\includegraphics[width=3.5in]{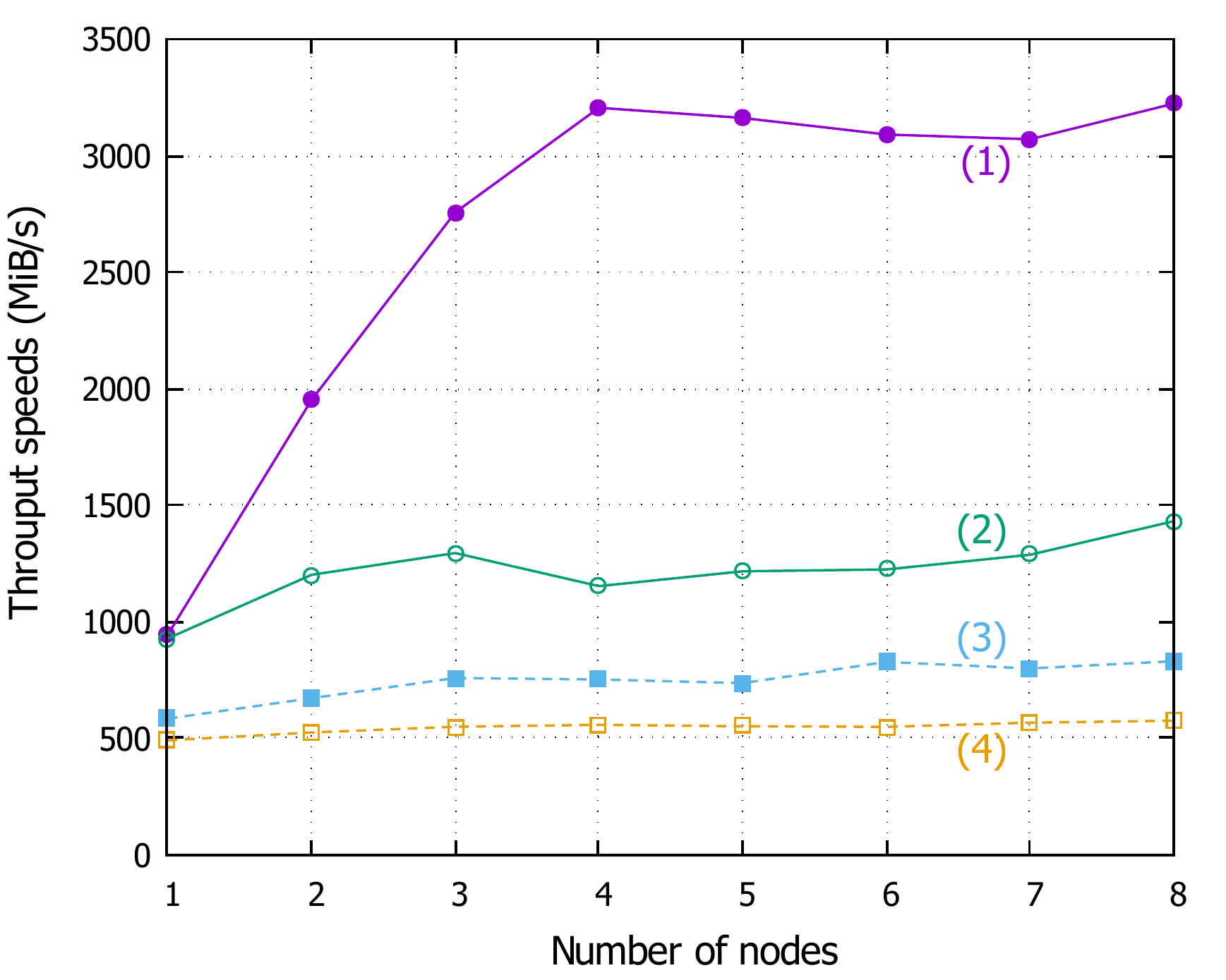}}
\caption{
The troughput between the front-end PCs and the event building PC.
The horizontal axis is the number of connections between the front-end PCs and the event building PC.
(1) shows the throughput without event building and recording.
(2) shows the throughput with event building but without recording.
(3) shows the throughput with event building and recording.
(4) shows the throughput with event building and compressed recording. 
}
\label{fig:eb_speeds}
\end{figure}

Figure.~\ref{fig:eb_speeds} shows the throughput of the measurements with several different conditions.
The horizontal axis means the number of data-transport connections.
Line (1) shows a throughput of the data transfer by multiple connections.
One connection can transfer data with up to 1.1~GiB/s.
However, the speed at a single connection did not achieve maximum speed by the overhead of buffer management.
The total data transfer throughput with over four connections achieved 3.1~GiB/s.
Line (2) shows the throughput with event building but without recording.
The throughput was limited to 1.2~GiB/s with over two connections.
The event building process filled up a CPU core and limited its performance.
The dotted line (3) shows the throughput with event building and with the data recording.
The recording process suppressed throughput to 0.80~GiB/s at over six connections case.
The throughput at under six connections became more slowly.
The recording process used a processor core fully, and the event building process had to wait for it.
The dotted line (4) shows the throughput with event building and with the compressed data recording.
The compression ratio of the data was 70~\% by the lz4 algorism.
The throughput went down a bit more and became 0.53~GiB/s.

\subsection{Performance evaluation of the local storage}
\begin{figure}[h]
\centerline{\includegraphics[width=3.5in]{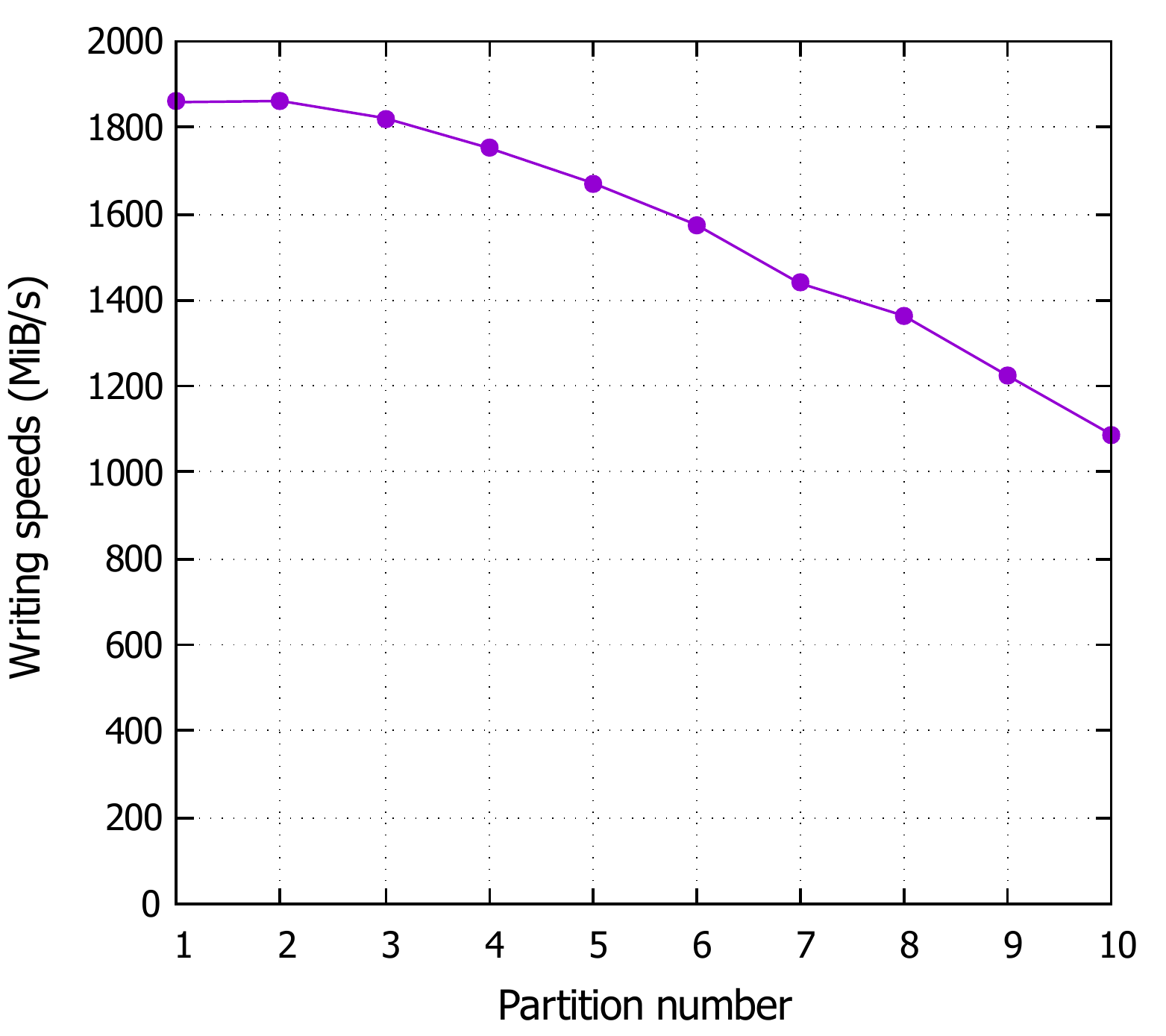}}
\caption{
The location dependency of the writing speed on the disk.
Partition one is most outside of the disk, and partition eight is inside of the disk.}
\label{fig:raid}
\end{figure}

The writing performances of the SAS RAID HDD system was measured.
The system is Broadcom/LSI MegaRAID SAS-3 and had ten SEAGATE 10~TB 7200~RPM hard disks.
The raw performance was measured by the standard "dd" command to compare major file systems "xfs" and "ext4".
The command option "oflag=direct" was given to ensure the measurements were not affected by the operating system's write buffer.
The writing speed of "xfs" was 1.84~GiB/s while 1.66~GiB/s for "ext4".
Thus, we adopted "xfs."
 Furthermore, hard disks have different writing speeds depending on the data location on the disk.
The location dependence of the writing speed is showing Fig.\ref{fig:raid} for ten divided hard disk partitions.

The writing speed of the slowest an innermost was at least 1.1~GiB/s.
The overall speed of writing was 1.5~GiB/s, which balanced the performance of the event building.

\subsection{Performance evaluation of the long-distance data transport}

\begin{figure}[h]
\centerline{\includegraphics[width=3.5in]{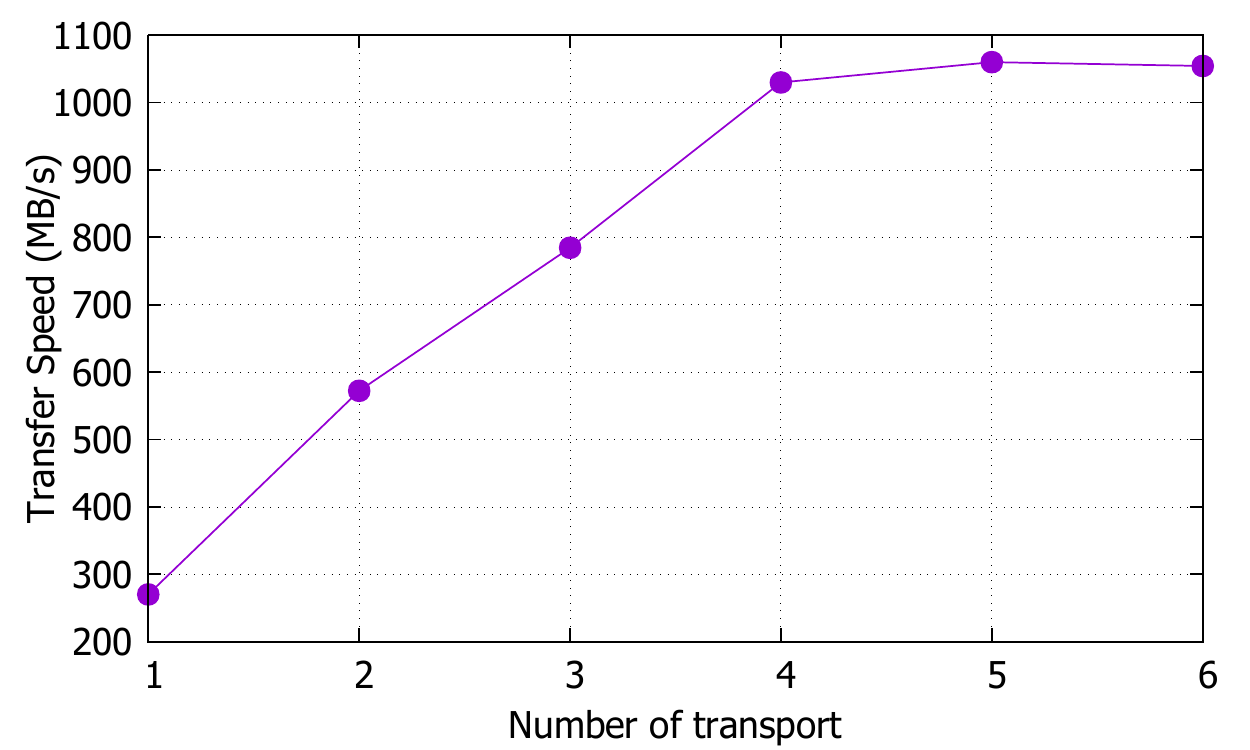}}
\caption{The total data copy speed by multiple "scp" comand. The horizontal axis means the number of "scp" command that runs siimultaneously.}
\label{fig:scp}
\end{figure}

The inter-laboratory data transfer throughput was confirmed using the common Linux "scp" command.
Figure~\ref{fig:scp} shows the data copy speed by multiple "scp" command.

The one "scp" command could copy data with 280 MiB/s.
The data copy's total speed achieved 1.0 GiB/s using over four "scp" process.

\section{Discussion of the entire performance of the data taking network}
The front-end network using the low price network switch worked well with over 300~MiB/s throughputs.
COMET Phase-I DAQ will use nearly ten front-end PC for the actual data taking.
And the back-end event building performance was limited to 1.2~GiB/s.
Therefore, the FS.COM network switch had good performance enough for the use in the front-end network.

The back-end network using the 10~Gbps Ethernet direct connection also has an adequate performance of 3.1~GiB/s with a high-speed multi-core processor since the multi-core processor can allocate one network interface handling per one core.
And each PCIe interface that has a NIC connects the processor directly and individually.
It means that is no interference outside of the processor, such as a shared bus architecture.
The speed of the direct connection is faster than in the case of 10~Gbps network switch.
The direct connection network also had an adequate performance.

The event building process can be a bottleneck since the whole data focus on it.
This is a structural week point for the tree structure network data taking.
The event building PC has a high-performance many-core server processor Xeon Gold 6126.
The processor is expected to process fast, has a wide bandwidth of memory access, and communicates quickly between the processor cores.
The event building process handled the data stream with 1.2~GiB/s throughput on the event building PC.

The intrinsic performance of the SAS RAID disk system is enough.
However, the recording process speed was a bottleneck in the entire system.
The recording process spent 100~\% calculation power of a processor core.
Therefore the throughput of the data taking decreased to 0.80~GB/s.

The speed of the long-distance data copy from J-PARC to KEK Tsukuba was confirmed that it has an acceptable bandwidth.
The RAID disk system can be separated into two by SCSI LUN.
And it has two SAS interface ports connected to the two back-end PCs (the event building PC and its backup).
The stored data can be expected to be transferred without interfering with the data taking by the double buffer method using the separated RAID disks and two back-end PCs. 

\section{Summary}

The COMET Phase-I DAQ is a two-layer network-based system.
The network of the DAQ assembled with the low price 1G/10G network switches for the front-end network and the switchless 10G network connection for the back-end network.
All components of the data taking network were evaluated, including the long-distance data copy.
The entire throughput of the data-taking is expected to achieve 0.80~GB/s, which is determined by the speed of the local data recording process.
It means that the data-taking system could be work 2.7 times faster than the estimated requirements.

\section{Acknowredgement}
The authors would like to express their gratitude to the COMET experimental group members.

\end{document}